\documentclass[pdflatex,sn-aps,iicol]{sn-jnl}% Basic Springer Nature Reference Style/Chemistry Reference Style
\usepackage{}
\usepackage{fancybox}
\usepackage{amsmath,amssymb,bm,amsthm}
\usepackage{graphicx}%
\usepackage{multirow}%
\usepackage{amsmath,amssymb,amsfonts}%
\usepackage{amsthm}%
\usepackage{mathrsfs}%
\usepackage[title]{appendix}%
\usepackage{xcolor}%
\usepackage{textcomp}%
\usepackage{manyfoot}%
\usepackage{booktabs}%
\usepackage{algorithm}%
\usepackage{algorithmicx}%
\usepackage{algpseudocode}%
\usepackage{listings}%
\def\beq{\begin{equation}}
\def\eeq{\end{equation}}

\def\<{\langle}
\def\>{\rangle}

\begin{document}

\title[Article Title]{A comparative review of recent results on supercritical anomalies 
in two-dimensional kinetic Ising and Blume-Capel ferromagnets}

\author*[1]{\fnm{Gloria M.} \sur{Buend\'ia}}\email{buendia@usb.ve}

\author[1,2]{\fnm{Celeste} \sur{Mendes}}\email{up202408487@edu.fc.up.pt}

\author*[3,4]{\fnm{Per Arne} \sur{Rikvold}}\email{p.a.rikvold@fys.uio.no}

\affil*[1]{\orgdiv{Department of Physics}, \orgname{Universidad Sim\'on Bol\'ivar}, \orgaddress{\city{Caracas}, \postcode{1080}, \country{Venezuela}}}

\affil[2]{\orgdiv{Faculdade de Ci\^encias }, \orgname{Universidade do Porto}, \orgaddress{\street{Rua do Campo Alegre}, \city{Porto}, \postcode{4169-007}, \country{Portugal}}}

\affil[3]{\orgdiv{PoreLab, NJORD Centre, Department of Physics}, \orgname{University of Oslo}, \orgaddress{\street{P.O. Box 1048 Blindern}, \city{Oslo}, \postcode{0316}, \country{Norway}}}

\affil[4]{\orgdiv{Department of Physics}, \orgname{Florida State University},\orgaddress{ \city{Tallahassee}, \state{Florida}, \postcode{32306-4350}, \country{USA}}}

\abstract{
Following the unexpected experimental discovery of ``sideband'' peaks in the fluctuation 
spectrum of thin Co films driven by a slowly oscillating magnetic field with a constant bias 
[P.~Riego et al., Phys.\ Rev.\ Lett.\ {\bf 118}, 117202 (2017)]
numerical studies of two-state Ising and three-state Blume-Capel (BC) ferromagnets in this 
dynamically supercritical regime have flourished 
and been successful in explaining this phenomenon. 
Here, we give a comparative review of this new literature and its connections to 
earlier work. Following an introduction and a presentation of the two models and 
the computational method used in many of these studies, we present numerical results 
for both models. Particular attention is paid to the fact that zero spins in the BC model 
tend to collect 
at the interfaces between regions of the two nonzero spin values, $\pm 1$. 
We present strong arguments that this phenomenon leads to a reduction of the effective 
interface tension in the BC model, compared to the Ising model.
}


%{keyword1, Keyword2, Keyword3, Keyword4}

\maketitle

%***********************************************************************NEW

\section{Introduction}
\label{sec:I}

Since the 1990's it has become clear that the well-known, equilibrium phase transition in the 
Ising universality class (e.g., a two-dimensional Ising ferromagnet in zero field at a critical 
temperature $T_c$) has a {\it nonequilibrium} analog. This dynamic phase transition, 
commonly known as DPT, occurs when a spin system in its ordered phase 
(i.e., below $T_c$) is perturbed by an oscillating magnetic field of period $P$. 
In this case, $P$ takes the role of $T$ in the equilibrium case, with a critical $P_c$ that 
depends on $T$ and the field amplitude $H_0$. 
For $P < P_c$, the magnetization oscillates near the {\it equilibrium} magnetization, while 
for $P>P_c$, the system follows the sign of the field, giving a 
constant, period-averaged magnetization, $\langle Q \rangle \approx 0$. 

Early studies were computational \cite{Tome1990,Lo1990,Rao1990,Chakrabarti1999}.
Kinetic Monte Carlo (MC) combined with finite-size scaling analysis
 \cite{Sides1998,Sides1999,Korniss2000,Robb2007,Buendia2008,Park2013}, as well as 
mean-field studies of Ising, Ginzburg-Landau, and $XY$ models 
\cite{Fujisaka2001,Gallardo2012,Idigoras2012,Robb2014,PAL24}, confirmed that this is 
a true DPT in the same 
universality class as the corresponding equilibrium Ising model. 
Experimental confirmations of Ising-like behavior in [Co/Pt]$_3$ magnetic multilayers 
\cite{Robb2008} and uniaxial Co films \cite{Berger2013} were later obtained. 

The close analogy in the critical region between the equilibrium Ising transition and the DPT
possibly delayed studies of the DPT in the {\it supercritical region}, $P > P_c$, by several years, 
and the experimental results presented in 2017 by Riego,
Vavassori, and Berger \cite{Riego2017} came as a surprise. 
In that work, Co films with $(10\underbar{1}0)$ crystallographic surface structure with a single,
in-plane magnetic easy axis were subjected to an oscillating
magnetic field plus a constant bias field, $H_b$. 
Such a constant bias field had previously been shown to be
a significant component of the field conjugate to $\langle Q \rangle$ in the
critical region near $P_c$ \cite{Robb2007,Robb2014,Gallardo2012,Idigoras2012,Berger2013}. 
In general, it favors spin transitions in the direction 
toward the state in which they have the same sign as $H_b$.
However, in stark contrast to the wide, unimodal maximum of the supercritical 
susceptibility of the equilibrium Ising model,
two distinct peaks in the corresponding response function 
were found at nonzero values of $H_b$, symmetrical about zero. 
These new results have inspired recent studies of DPT transitions by effective-field \cite{SHI19} 
and Monte Carlo methods \cite{YUKS21}, and 
also in the presence 
of nonantisymmetric and random fields and interactions \cite{QUIN24,LI24,YUKS24,VATZ24}. 

In this short paper, we review central results from recent, kinetic MC studies of the 
2D Ising ferromagnet \cite{Buendia2017} and BC model \cite{MEND24}, both in the dynamically 
supercritical regime, $P > P_c$. 
The defining Hamiltonians and the MC simulation method used are presented in Sec.~\ref{sec:H},
while numerical results for the Ising model and the BC model 
are presented in Sec.~\ref{sec:IS} and Sec.~\ref{sec:BC}, respectively. 
Some nucleation-theoretical results relevant for both models are discussed in Sec.~\ref{sec:Nuc}.
Conclusions and a discussion are given in Sec.~\ref{sec:Conc}.

\section{Models and simulation method}
\label{sec:H}
We consider kinetic spin models with a time-dependent external
field $H(t)$ plus a constant ``bias field'' $H_b$, 
ferromagnetic nearest-neighbor interactions, $J > 0$, 
and a crystal field, $D$.  The defining Hamiltonian is
\begin{equation}
{\mathcal H}=-J\sum_{\langle ij \rangle} s_{i}s_{j} - D\sum_{i} s_{i}^2  -  \left[ H(t)+H_{b} \right] \sum_{i} s_{i} \;,
\label{eq=ham}
\end{equation}
where the first sum runs over all nearest-neighbor pairs,
and the other two over all sites. $H(t)$ is a symmetrically oscillating external field of period $P$,
\begin{equation}
 H(t)=H_{0}  \cos \left( \frac{2 \pi}{P} t \right) \;.
\label{eq:field}
\end{equation}
$H_0$, $H_{b}$, $D$  are all given in units of  
$J$ (i.e., $J=k_{\rm B}=1$), where $k_{\rm B}$ is
Boltzmann's constant. The systems are simulated on a square lattice of $N=L \times L$ sites with
periodic boundary conditions at constant temperature, $T = 0.8 T_c$.
If $s_{i}=\pm1$, the $D$ term becomes an unimportant constant,  and 
the above Hamiltonian describes the standard Ising model. However, 
with $s_{i}\in [ \pm1,0 ]$ and $D \in [ - \infty, + \infty ] $ it describes the BC model. 
(In the limit $D = + \infty$, $s_i = 0$ becomes forbidden, and the BC model degenerates to the 
two-state Ising model.) 
Kinetic 
MC simulations were performed with a time unit of one MC step per site (MCSS), during which 
each site was visited once on average. $P$ is in units of MCSS. 
Transition probabilities were given by a heat-bath single-spin transition dynamics. 
A spin $s_i$ was selected at random and changed to $s_i'$  with probability
\begin{equation}
 W(s_{i}\rightarrow s_{i}')
 =\frac {\exp[-\beta \Delta E(s_{i},s_{i}')]}{\sum_{s_i'}\exp[-\beta \Delta E(s_{i},s_{i}')]} \;,
\label{eq:W}
\end{equation}
where $\Delta E(s_{i},s_{i}')$ is the change in the system energy associated
with changing the spin $i$ from $s_i$ to $s_i'$, and $\beta=1/k_{\rm B}T$. 
$s_i$ and $s_i'$ are restricted to the allowed values for each model: $\pm 1$ for Ising and 
$[ \pm 1, 0]$ for BC. 

Quantities calculated were the time dependent, normalized magnetization per site,
\begin{equation}
 m(t)=\frac{1}{L^2} \sum_{i}s_{i}(t) 
 \label{eq:mag}
\end{equation}
and its integrals over each field cycle $k$, 
\begin{equation}
 Q_{k}=\frac{1}{P}  \int_{(k-1)P}^{kP} m(t)dt \;.
 \label{eq:q}
\end{equation}
The dynamic order parameter
is the period-averaged magnetization, $\langle Q \rangle$, 
defined as the average of $Q_k$ over many periods. Its fluctuations are measured by the scaled variance,
\begin{equation}
\chi_{L}^{Q}=L^{2}(\langle Q^{2} \rangle -\langle Q \rangle^{2}) \;.
\label{eq:var}
\end{equation}
Most measurements were taken over 800 field periods, after discarding 200 periods. 
Thus, at least $800 \times P$ MCSS in the supercritical region with $P > P_c$ 
were performed for each measurement.

\section{Kinetic Ising model}
\label{sec:IS}

We first present results for the standard Ising model \cite{Buendia2017}, 
Eq.~(\ref{eq=ham}) with $s_{i}=\pm1$. 
Most calculations were performed with  $H_{0}=0.3$ at $T=0.8T_{c}$, where
$T_{c}={2}/{\ln(1+\sqrt {2})} \approx 2.269$ is the critical temperature of the standard,
square-lattice Ising model in zero field. The  critical
period of this oscillating field with zero bias is $P_c \approx 258$ \cite{Sides1999}. 

\begin{figure}[ht]
\begin{center}
\vspace*{-0.8truecm}
\includegraphics[angle=0,width=.48\textwidth]{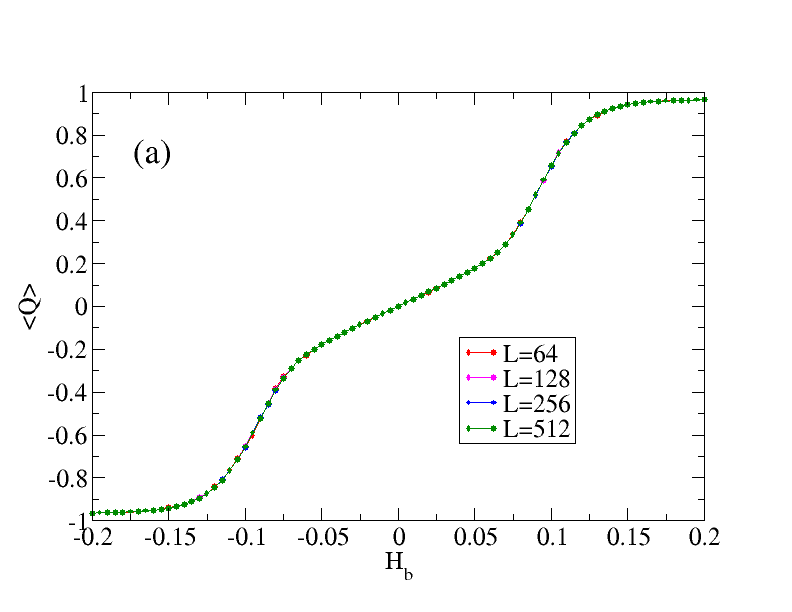}
\end{center}
\begin{center}
\vspace*{-0.8truecm}
\includegraphics[angle=0,width=.48\textwidth]{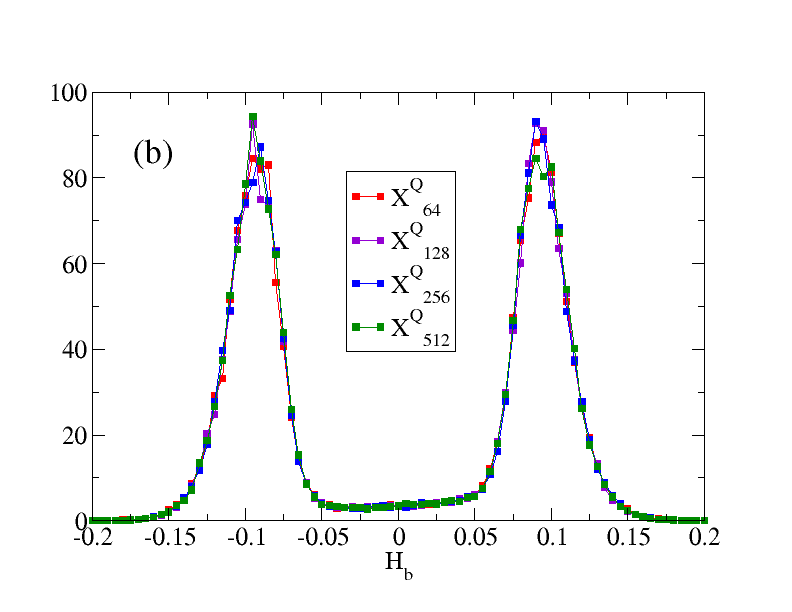}
\end{center}
\begin{center}
\vspace*{-0.3truecm}
\caption[]{
\baselineskip=0.15truecm
Plots of the order parameter $\langle Q \rangle$ (a), and the scaled variance $\chi_{L}^{Q}$ (b)
vs $H_b$ for the kinetic Ising model.
$P = 1000 \, {\rm MCSS} \approx 3.9P_c$. 
For this range of sizes, we cannot detect any finite-size effects.  
Error bars are smaller than the symbol size.
See discussion of this figure in Sec.\ \protect\ref{sec:IS}.
Figures adapted from Ref.~\cite{Buendia2017} with permission.
}
\end{center}
\label{fig:size1}
\end{figure}

In Fig.\ \ref{fig:size1} we present results for $P  \approx 3.9P_c$ 
and for several lattice sizes. Figure~\ref{fig:size1}(a) shows $\langle Q \rangle$ vs  $H_b$. 
For weak $H_b$, $\langle Q \rangle$ increases almost linearly until, at $|H_b | \approx 0.09$, the 
slope increases and finally saturates for $|H_b| \gtrsim 0.15$. 
This behavior is reflected in the scaled variance $\chi_L^Q$ shown in Fig.\ \ref{fig:size1}(b), 
which presents two symmetrical fluctuation peaks located at  
$|H_b|=|H_b^{\rm peak}| \approx 0.09$,
 separated by a flat-bottomed valley and saturating for strong bias. 
For these values of $L$ and $P$ we cannot see any finite-size effects that could suggest that 
this is a critical behavior associated with a phase transition. The appearance of these 
``sidebands'' is consistent with the experimental results presented in Ref.\ \cite{Riego2017} 
and clearly demonstrates that they are not due to any residual magnetostatic 
long-range interactions.

\begin{figure}[ht]
\begin{center}
\vspace*{-2.0truecm}
\includegraphics[angle=0,width=.48\textwidth]{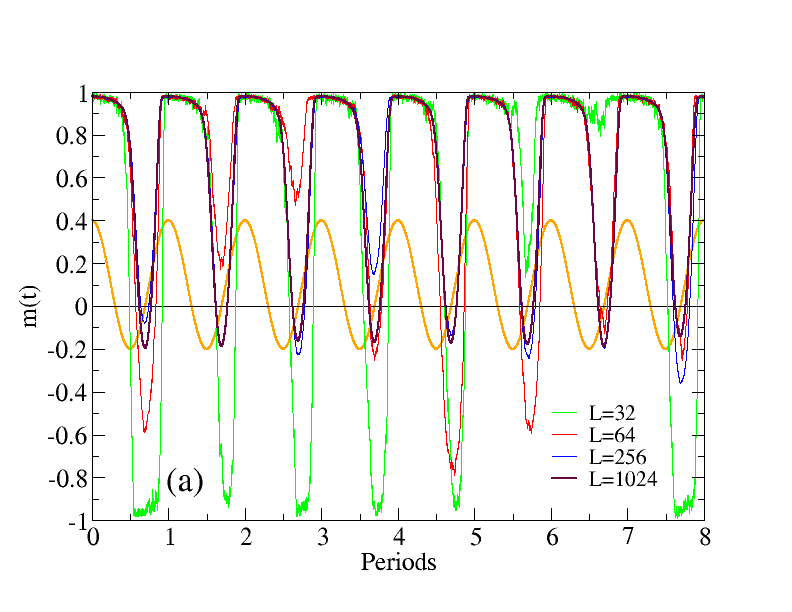}
\end{center}
\begin{center}
\vspace*{-1.2truecm}
\includegraphics[angle=0,width=.48\textwidth]{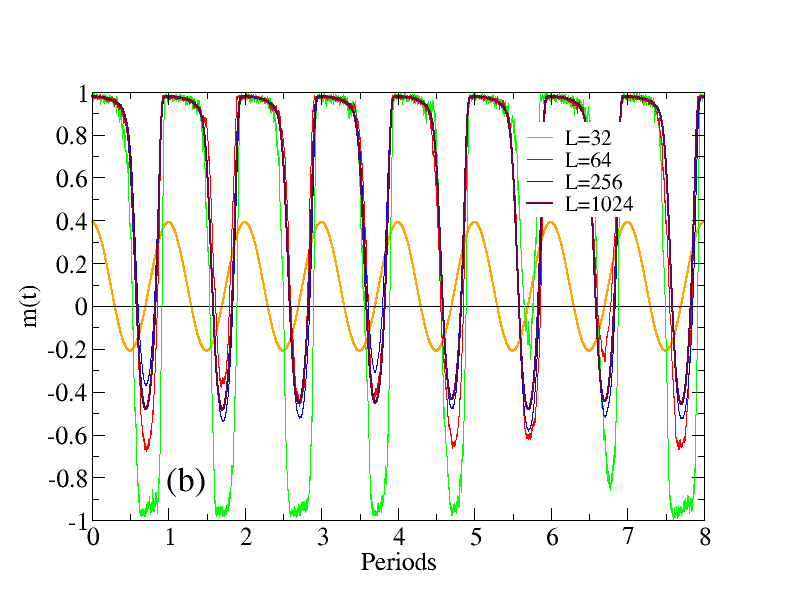}
\end{center}
\begin{center}
\vspace*{-1.2truecm}
\includegraphics[angle=0,width=.48\textwidth]{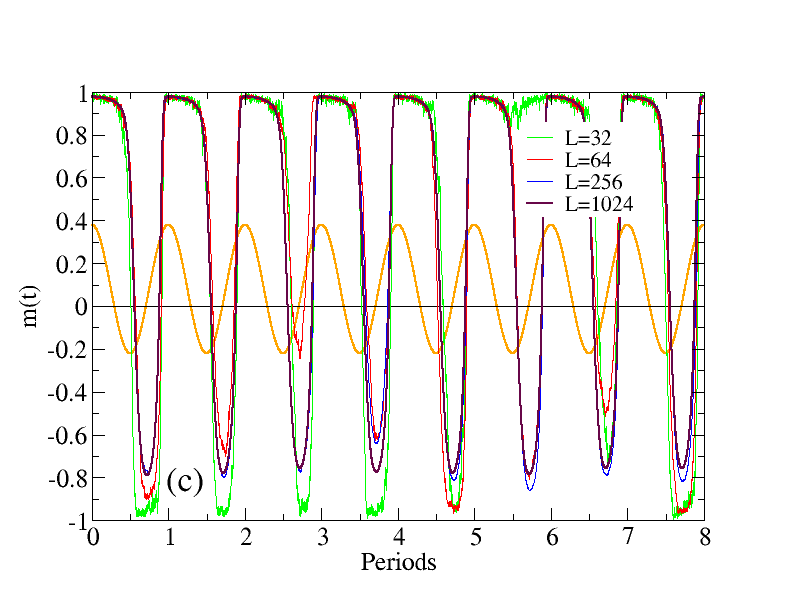}\end{center}
\vspace*{-0.3truecm}
\caption[]{
\baselineskip=0.15truecm
Time dependent magnetization $m(t)$ over eight cycles following a $200P$ stabilization run 
 for the kinetic Ising model. 
The orange curves are the total applied field, $H(t) + H_b$. The bias is always positive, and 
$P=1000$ MCSS. (a) $H_b=+0.10$, just on the strong-bias side of the fluctuation peak (b)  
$H_b=+0.0915$, near the maximum of the fluctuation peak. (c) $H_b=+0.08$, 
just on the weak-bias side of the peak. 
The switching is stochastic for the smaller systems (SD regime), 
becoming more deterministic with increasing $L$ ( MD regime). 
Figures adapted from Ref.~\cite{Buendia2017}  with permission.
}
\label{fig:tenP}
\end{figure}

To explore the nature of the ``sidebands'' and their dependence on the system size  
and bias, in Fig.\ \ref{fig:tenP} we present short time series of the normalized magnetization, 
$m(t)$. Since $H_b > 0$, the up-spin phase is favored, while the down-spin phase is disfavored.
 Figure \ref{fig:tenP}(a) was obtained on the strong-bias side. Size effects are notable in this case. 
 For $L=32$ and 64 the switching from the favored to the disfavored magnetization is stochastic 
 and abrupt. For the larger systems, the switching becomes more deterministic and gradual, but 
 the spins do not have time to reach the complete down phase before the field switches 
 to positive values. Figure \ref{fig:tenP}(b) was obtained at $H_b\approx H_b^{\rm peak}$. 
 While the switching remains stochastic for
  $L=32$, the larger systems have a more deterministic behavior 
  and reach larger negative magnetizations than in the previous case. 
  Figure \ref{fig:tenP}(c) was obtained at $H_b$ on the weak-bias side. The switching for $L=32$ remains stochastic, but all the other sizes have a deterministic behavior and perform a more
   complete switching toward the negative phase.

\begin{figure}[ht]
\begin{center}
\includegraphics[angle=0,width=0.48\textwidth]{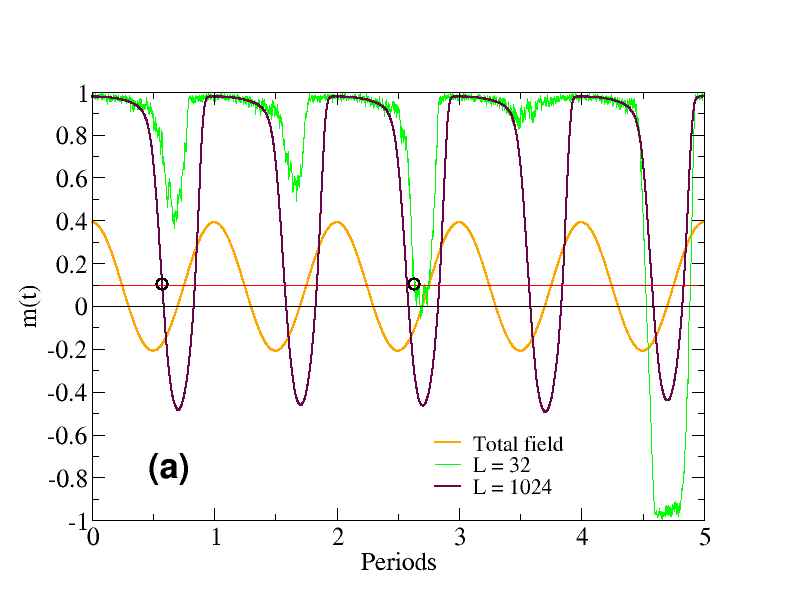}
\end{center}
\begin{center}
\vspace*{-0.8truecm}
\includegraphics[angle=0,width=0.45\textwidth]{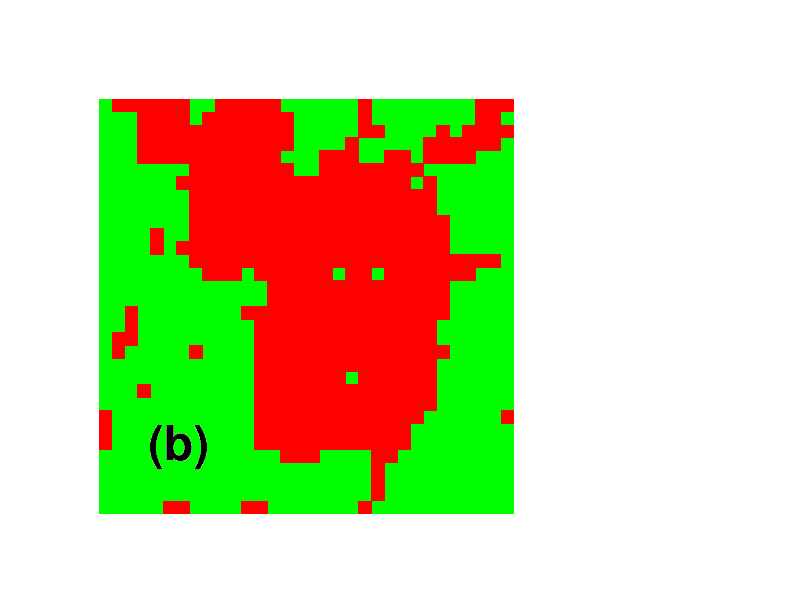} 
\end{center}
\begin{center}
\vspace*{-1.7truecm}
\includegraphics[angle=0,width=0.45\textwidth]{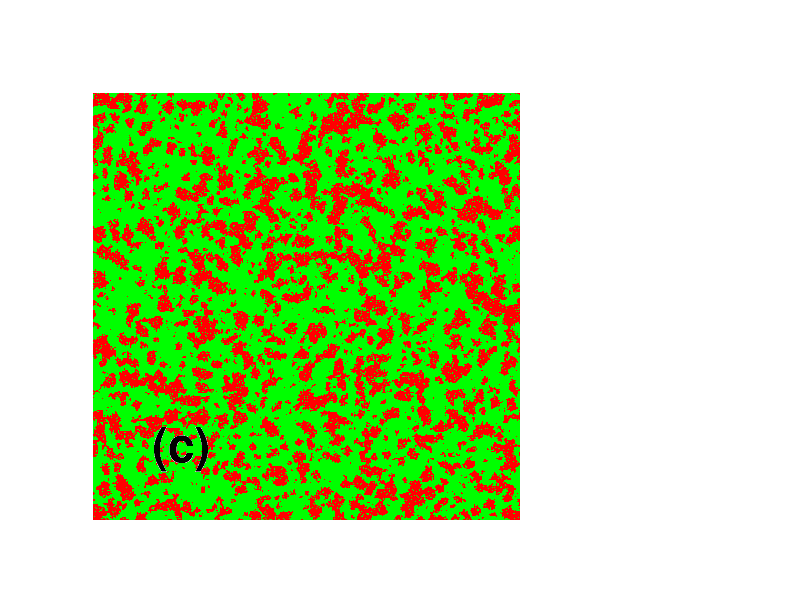} 
\end{center}
\vspace*{-1.3truecm}
\caption[]{
\baselineskip=0.15truecm
A short time series and snapshots showing the growth of disfavored-phase clusters 
in the kinetic Ising model for 
$P=1000$ MCSS at $H_b = +0.0915$, near the corresponding peak position.
(a)
Time series $m(t)$ over five periods, following $200P$ stabilization, 
for $L=32$ (green) and 1024 (maroon).  
The total applied field, $H(t) + H_b$, is shown in orange.
The snapshots were captured the first time that 
$m(t)$ fell below $+0.1$ (red horizontal line), 
corresponding to a disfavored-phase (down-spin) fraction of $0.45$. The times of capture are 
marked by black circles. 
In the snapshots, up spins are green, and down spins are red. 
(b)
$L=32$. A single droplet of the down-spin phase has just nucleated. 
The time is approximately 0.13 periods past 
the third minimum of the total applied field, as seen in part (a). 
(c)
$L=1024$. Many down-spin droplets have nucleated 
at different times and then grown almost 
independently. At the moment of capture, some clusters have coalesced while others are still 
growing independently. From part (a) it is seen that this MD 
switching mode leads to a nearly deterministic evolution of $m(t)$, reaching a down-spin fraction 
of 0.45 reliably approximately 0.07 periods past each minimum of the total applied field.
Figures adapted from Ref.~\cite{Buendia2017}  with permission.
}
\label{fig:SNAP}
\end{figure}

To further illustrate the different decay mechanisms in Fig.~\ref{fig:tenP}(b), 
in Fig.~\ref{fig:SNAP}(a) we present time 
 series for $m(t)$ over several periods with $P=1000$ MCSS and a bias 
 $H_b = +0.0915$, near the peak of $\chi_L^Q$ with $L=32$ and 1024. 
 Snapshots taken at $m(t) = +0.1$ 
(indicated by circles in the figure), when the total applied field is negative, show the nucleation 
  and growth of droplets of the stable, disfavored spin-down phase. 
Figure \ref{fig:SNAP}(b), for $L=32$, is a snapshot showing the growth of a single droplet of 
the stable down phase (SD regime), 
while Fig.~\ref{fig:SNAP}(c) shows that for $L=1024$ the decay 
occurs by the independent nucleation, growth, and eventual 
coalescence of many droplets of the stable phase. 
This multi-droplet (MD) decay leads to a deterministic regime that is well described by the
Kolmogorov-Johnson-Mehl-Avrami (KJMA) approximation 
\cite{Rikvold1994,BIND16,Kolmogorov1937,Johnson1939,Avrami1_1939,Avrami2_1939,Avrami3_1939,RAMO99}.

These results clearly indicate that the ``sidebands'' are a result of a change of the switching 
mechanism. In  the vicinity of the peaks, a cross-over from a stochastic single-droplet mechanism 
 (for the smaller systems or strong bias) to a nearly deterministic multidroplet mechanism 
 (for the larger system or weak bias), in agreement with known results for field-driven 
magnetization  switching by homogeneous nucleation and growth of droplets of the 
stable phase \cite{Rikvold1994}.

\section{Kinetic Blume-Capel model}
\label{sec:BC}

In this section we present results for the ferromagnetic kinetic BC model \cite{MEND24}, 
defined by Eq.~(\ref{eq=ham})
 with $s_i \in [ \pm 1, 0 ]$, and $D \in [ -1, +1 ]$.  This system has a more complex 
 equilibrium phase diagram than the Ising model 
 \cite{Lawrie_Sarbach_1984, Per1988, Wilding1996, Selke_2010, Buendia2023}. 
 For high temperatures, zero magnetic field, and greater values of $D$, it undergoes a continuous 
 (second-order) equilibrium transition between a ferromagnetic and a paramagnetic phase, 
 while for low 
 temperatures and sufficiently negative values of $D$, the transition is discontinuous (first-order) 
 \cite{Blume1971, JGBrankov_1972, PhysRevB.15.1602, Ng1978, BALCERZAK200487, Zaim2008}. 
 The two transition lines join smoothly at a tricritical point ($D_t, T_t$).
\cite{Kwak2015}.  (See Fig.~1 of Ref.~\cite{MEND24}.) 

For values of $D$ and $T<T_c$, corresponding to the continuous equilibrium 
transition, the BC model 
in an oscillating magnetic field of period $P$ has a DPT at a critical period $P_c$, 
analogous to the standard Ising model discussed in Sec.~\ref{sec:IS} 
above. This dynamic transition also belongs to the same universality class as the 
equilibrium Ising model \cite{Vatansever2018}. In this section we show that the BC model also 
presents ``metamagnetic" anomalies in the supercritical region, $T<T_c$, $P>P_c$, 
just like the Ising model.

Heat-bath MC simulations as defined in Eq.~(\ref{eq:W}) were again performed on a 
square lattice of linear dimension $L$ with periodic boundary conditions, 
at $T = 0.8T_c$ and with $H_0=0.2$.  (Note that $H_0$ is different from the value used in the 
simulations of the standard Ising model above.)
For the BC model, $T_c$ depends on $D$, 
and $P_c$ depends on both $D$ and $H_0$. In Table \ref{tablapcritico} we report the critical 
temperatures calculated by Malakis et al. \cite{Malakis2010} and the critical periods calculated by 
Mendes et al.\ \cite{MEND24} for $H_0=0.2$. The data were obtained for values of $D$ within the 
region where the phase transition is continuous, and in the supercritical region where 
$P(D)>P_{c}(D)$, 
a region where, as with the Ising model,  we expect to find the metamagnetic anomalies. 
Indeed, data for $\langle Q \rangle$ and $\chi^Q_L$ for different $L$ are only 
weakly size dependent, qualitatively similar to those for the 
Ising model. Detailed results are found in Ref.~\cite{MEND24}.

\begin{table}[t]
    \centering
    \caption{Critical periods $P_c$, peak positions $|H_b^{\rm peak}|$, and 
    critical temperatures $T_c$, for the kinetic BC model 
    subjected to a sinusoidal field of amplitude $H_0 = 0.2$ for different values of crystal field 
    $D$ at $T = 0.8 T_c$. The data for $T_c$ were obtained from \cite{Malakis2010}, 
    and $P_c$ and $|H_b^{\rm peak}|$ from \cite{MEND24} under CC-BY 4.0 license.}  
    \begin{tabular}{|c|c|c|c|}
          \hline
         $D$ & $P_c$ (MCSS) & $|H_b^{\rm peak}|$ & $T_c$\\
         \hline
         0 & 870 $\pm$ 10 & 0.060 $\pm$ 0.001 & 1.693(3)\\
         \hline
         $-0.5$ & 720 $\pm$ 10 & 0.061 $\pm$ 0.002 & 1.564(3)\\
         \hline
         $-1.0$ & 460 $\pm$ 10 & 0.062 $\pm$ 0.002 & 1.398(2)\\
         \hline
        $-1.5$ & 205 $\pm$ 5 & 0.063 $\pm$ 0.002 & 1.151(1)\\
         \hline
         $-1.75$ & 115 $\pm$ 5 & 0.067 $\pm$ 0.002 & 0.958(1)\\
         \hline
    \end{tabular}
    \label{tablapcritico}
\end{table}

\begin{figure}[ht]
\begin{center}
    \includegraphics[width=0.48\textwidth]{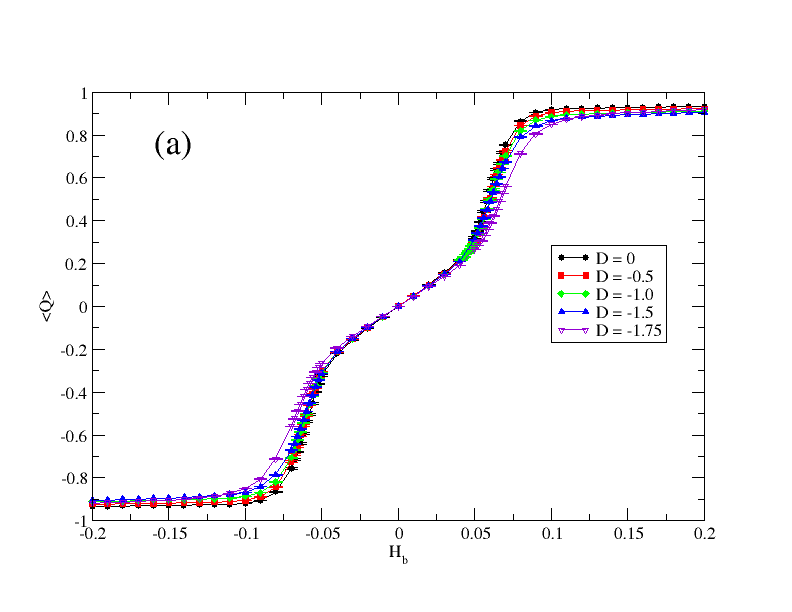}
\end{center}
\begin{center}
    \includegraphics[width=0.48\textwidth]{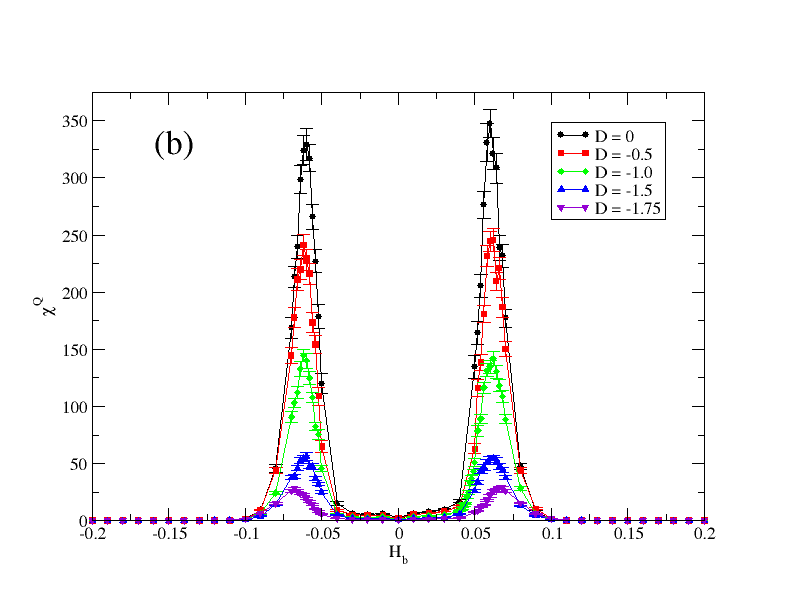}
%\hspace{20pt}
\end{center}
     \caption[]{
     \baselineskip=0.15truecm
     Dependence of the sidebands  for the kinetic BC model on the crystal field $D$ 
     for $L = 128$. Here, $D \in[-1.75, 0]$, so zeros are significant. 
     For each value of $D$, we maintained $P \approx 4 P_c(D)$ and 
     $T = 0.8 T_c(D)$. The values of $P_c(D)$, $| H_b^{\rm peak} |(D)$, and $T_c(D)$ 
     are given in Table \ref{tablapcritico}. 
     (a) $\langle Q \rangle$ vs $H_b$. (b) $\chi^Q$ vs $H_b$. 
As $D$ decreases toward $D_t$, the peak height decreases significantly 
while $|H^{\rm peak}_b|$ increases slightly. 
Figures from Ref.~\cite{MEND24} under CC-BY 4.0 license.
     }
\label{distintosD}
\end{figure}

As the crystal field $D$ approaches its tricritical value 
from the positive side, the density of zeros increases. Figure~\ref{distintosD} shows the results. 
The order parameter in Fig.~\ref{distintosD}(a) exhibits a strong dependence on $D$ only 
near $|H^{\rm peak}_b|$. 
The slope of $\langle Q \rangle$ around $|H^{\rm peak}_b|$ decreases as $|D - D_t|$ 
is decreased. This is reflected in the decreasing heights of the scaled variance peak in 
Fig.~\ref{distintosD}(b). The results also suggest that $|H^{\rm peak}_b|$ increases as 
$D$ closely approaches $D_t$. This is shown in 
Table~\ref{tablapcritico}. These results indicate that, as the density of zeros increases, 
stronger bias fields are needed to generate the asymmetry responsible for
the sidebands, and even for these larger values of $H_b$, the heights of the fluctuation 
peaks are reduced. This is consistent with the assertion by Schick and Shih  \cite{Schick1986}, 
that the effective interface tension should decrease in proportion to $|D - D_t|$ close to the
tricritical point. 

From a very detailed study of the dependence of the sidebands on the period $P$, the cystal field
 $D$, and the amplitude of the oscillating field $H_0$, it was concluded 
 that their behavior is analogous to that of the Ising model \cite{MEND24}. 
 In the BC model, as the density of zeros 
 is increased by letting $D$ approach $D_t$, larger bias fields are 
 needed to generate the asymmetry responsible for the sidebands. This results in flatter peaks 
 \cite{MEND24}.

\begin{figure}[ht]
\begin{center}
    \includegraphics[width=0.48\textwidth]{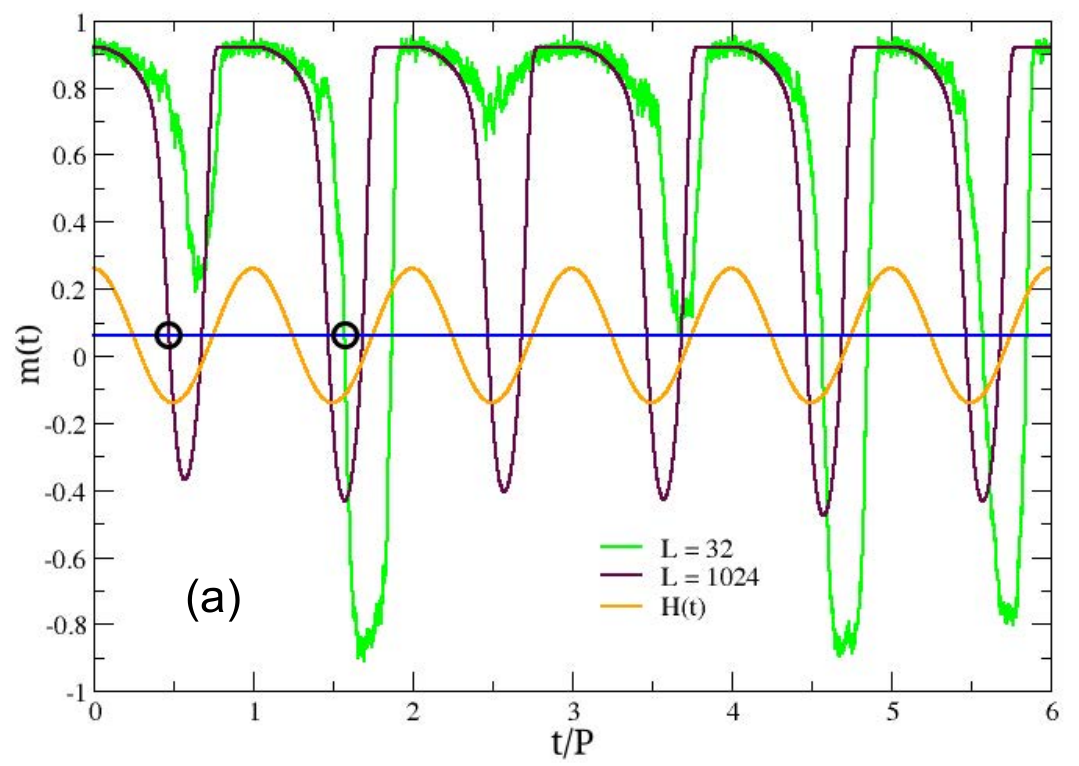}
    \end{center}
    \begin{center}
\vspace*{-10pt}
    \includegraphics[width=0.25\textwidth]{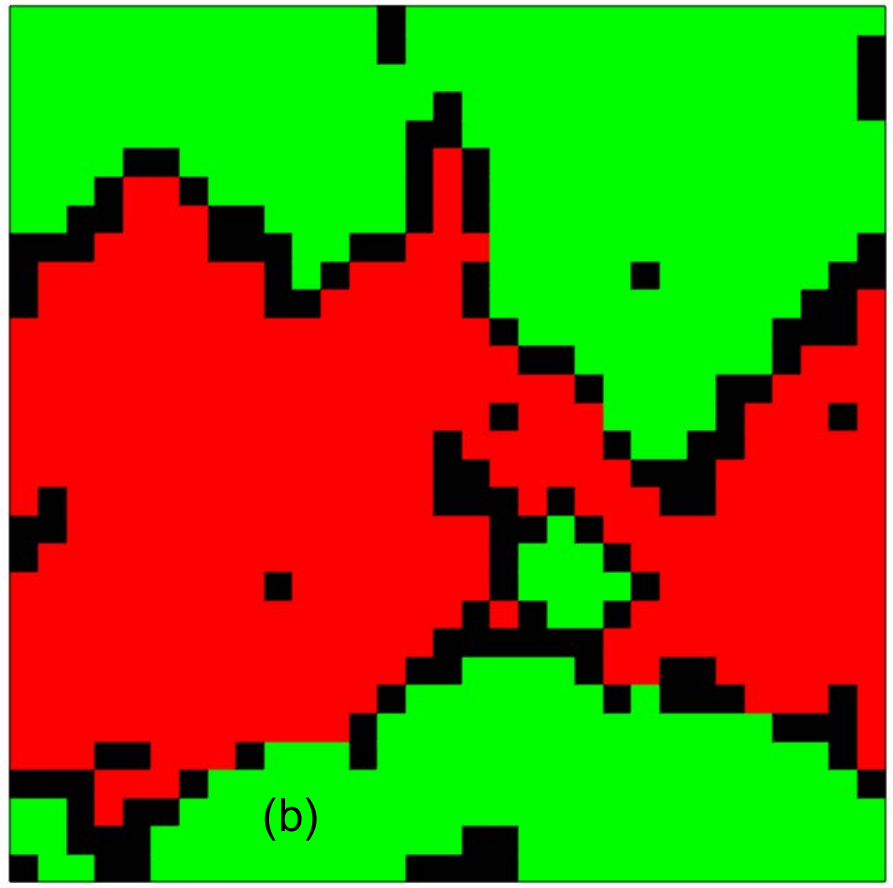}
    \end{center}
    \begin{center}
    \vspace*{+1.0truecm}
    \includegraphics[width=0.25\textwidth]{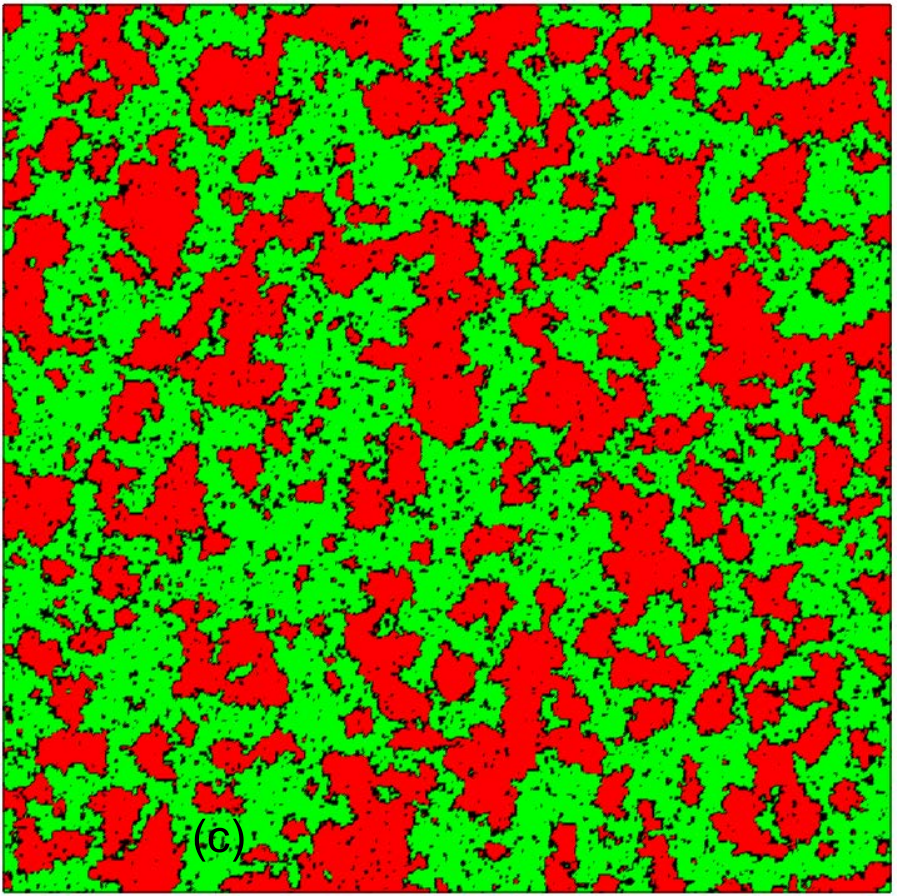}
        \end{center}
     \caption[]{
          \baselineskip=0.15truecm 
          Results for the kinetic BC model. 
          (a) $m(t)$ over six periods for $L = 32$ and $L = 1024$. 
     The total applied field $H(t) + H_b$, with $D = -1.5$, $P = 820 \approx 4 P_c$ and 
     $H_b = 0.063 \approx |H^{\rm peak}_b|$, is shown in orange. Snapshots of the lattice 
     were taken for both sizes when $m(t)$ first reached the threshold of 0.063 
     (blue horizontal line). The times of capture are marked by black circles. 
     For this threshold value, $H(t) + H_b < 0$, so the disfavored configuration of down-spins 
     is the stable state, which is shown in red. Up-spins 
     are shown in green, and zeros in black. (b) Snapshot for $L = 32$. 
     The decay occurs by the SD mechanism. 
     (c) Snapshot for $L = 1024$. The decay occurs by the MD mechanism. 
     In both snapshots, the zeros accumulate at the cluster boundaries. 
     Figures reproduced from Ref.~\cite{MEND24} under CC-BY 4.0 license.
     }
        \label{snapshot}
\end{figure}

As with the Ising model, we compare the magnetization time series for different system sizes. 
Stochastic behavior for $L=32$ and nearly deterministic behavior for $L=1024$ are 
shown in Fig.~\ref{snapshot}(a). 

In Fig.~\ref{snapshot}(b) and (c) we present snapshots of the system at a point where it 
is in the process of going from the metastable phase to the stable one. 
It is clear that for the small system, Fig.~\ref{snapshot}(b), this process occurs by the growth 
of a single droplet of the stable  phase (SD), while for the larger system, Fig.~\ref{snapshot}(c), 
the decay to the stable phase occurs by the independent nucleation and growth of many droplets 
(MD).  This behavior, as in the Ising model, is consistent with nucleation theory.

From Fig.~\ref{snapshot}(b) and (c) we also notice that, in both regimes, SD and MD,  the 
zero spins are mostly located at the interfaces between the two nonzero phases. 
This may result in a reduced interface tension for the BC model,  compared to the 
Ising model \cite{Schick1986}.

\section {Nucleation theory for the kinetic spin models}
\label{sec:Nuc}

\begin{figure}[ht]
\begin{center}
\vspace*{0.8truecm}
\includegraphics[angle=0,width=.5\textwidth]{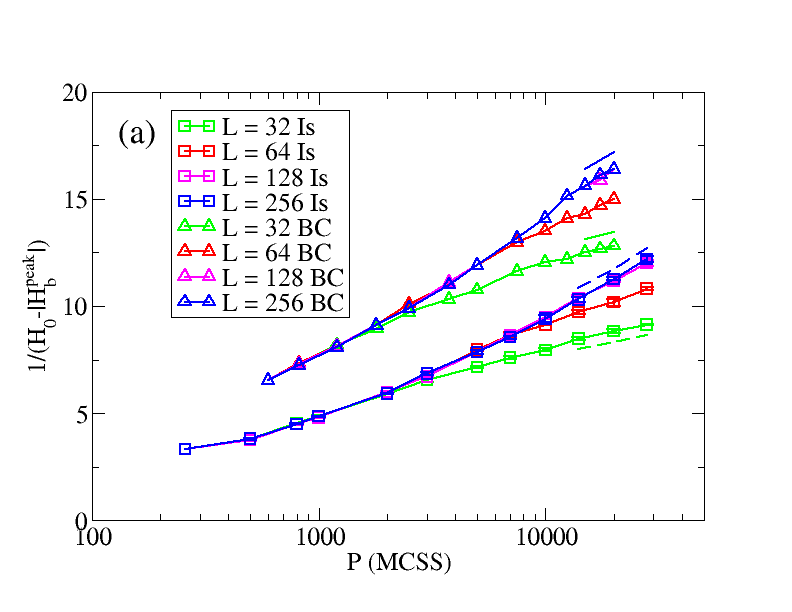} 
%\hspace*{0.4truecm}
\includegraphics[angle=0,width=.5\textwidth]{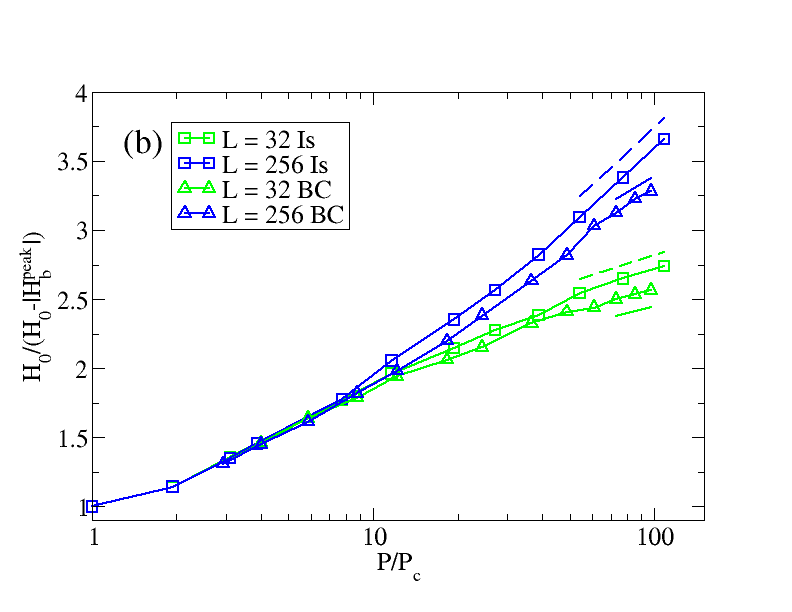} 
\end{center}
\vspace*{-0.3truecm}
\caption[]{
\baselineskip=0.15truecm
Linear vs logarithmic plots of peak positions vs oscillation period, corresponding to 
the nucleation-theoretical relation, Eq.~(\ref{eq:pekreq1}). 
(a)
Data for the Ising model \cite{Buendia2017} and the BC model \cite{MEND24}, displayed 
together. As the simulations were performed with different values of $H_0$, 0.3 and 0.2, 
 respectively, the results do not coincide. 
 (b)
 The same data, plotted in terms of the dimensionless variables, $1/(1 - H_b^{\rm peak} / H_0)$ 
 and $P/P_c(H_0)$. The scaled data coincide for $P/P_c < 10$. 
 For $P/P_c > 10$, results for the different models and different system sizes separate, 
 as predicted by the theory. 
 See further discussion in 
 Sec.~\ref{sec:Nuc}. 
Data obtained from Refs.~\cite{Buendia2017} with permission and from 
\cite{MEND24} under CC-BY 4.0 license.
}
\label{fig:PPos}
\end{figure}

By studying the dependence of $|H^{\rm peak}_b|$ on the period length of the 
external oscillating field 
and employing nucleation theory on the kinetic models subjected to a sinusoidal field 
plus a constant bias, the following relation was expected to apply 
for very large periods \cite{Buendia2017, MEND24}
\begin{equation}
P \sim L^{-a} \exp \left( \frac{1}{b} \frac{\Xi_0}{H_0 - |H_b^{\rm peak}| } \right ),
\label{eq:pekreq1}
\end{equation}
with $a=2$ and $b=1$ for SD switching, and $a=0$ and $b=3$ for
MD switching, where  $\Xi_{0}$ is the field-independent part of the free-energy cost of a 
critical droplet divided by $k_{B}T$. For details  see \cite{Buendia2017} and references therein.
From this expression we expect that, if we plot $\frac{1}{(H_0 - |H^{\rm peak}_b|)}$ vs $\log P$, 
we will get a straight line for large $P$, such that the ratio between the slopes for the MD 
(large system) and SD (small system) regime  for the same model is 3. Such plots 
for both the Ising and BC models are shown together in Fig.~\ref{fig:PPos}(a). 
For the Ising model, the ratio of the slopes is approximately 2.866(1) \cite{Buendia2017}, 
and for the BC model the ratio is approximately 2.405(1) \cite{MEND24}. 
Both values, particularly the one for the Ising model, are close to the expected value of 3. 
A work based on effective field theory on the kinetic BC model in the presence of a bias 
reported a slope of 2.236 \cite{Yuksel2022}. 

The results for the two models that are shown in Fig.~\ref{fig:PPos}(a) 
were obtained from simulations 
with different values of $H_0$, which leads to different values for $P_c$. In order to directly 
compare the results, in Fig.~\ref{fig:PPos}(b) we present them in terms of 
dimensionless variables, 
$1/(1 - H_b^{\rm peak}/H_0)$ vs $P/P_c(H_0)$. 
With this scaling, the curves collapse for $P/P_c < 10$. 
For $P/P_c >10$, the difference between the two models
due to the zeros in the BC case is evident. 
In each case the behavior is consistent with the predictions of nucleation theory.
From this, we can also obtain a rough estimate for the ratios of $\Xi_0$ (which 
is proportional to the square of the effective interface tensions \cite{RIKV94}) 
for the same system size 
for the two models. For $L=256$ we obtain $\Xi_0 ({\rm BC}) / \Xi_0 ({\rm Is}) 
\approx 0.658(1)$ and for $L=32$ we get $\approx 0.784(1)$.  
The fact that both ratios are less than unity 
strengthens the case that the zero state in the BC model acts as a surfactant to 
reduce the effective interface tension between the two nonzero phases  
\cite{Schick1986,CIRI96,CIRI24}. 
A very recent study indicates that the surface-tension scaling of the BC model on a 
triangular lattice agrees with what has previously been observed for the 
square-lattice \cite{MATA25}.

\section{Conclusions}
\label{sec:Conc}

The experimental discovery of ``sideband'' peaks in the fluctuation 
spectrum of thin Co films driven by a slowly oscillating magnetic field with a constant bias 
\cite{Riego2017}
was quickly followed by a number numerical studies (see the list of references) 
that provided successful explanations of this nonequilibrium phenomenon. 
In the comparative study presented here, we show that the supercritical anomalies in the 
two-state Ising and the three-state BC model share many common features. These include a 
gradual crossover between SD switching behavior for smaller systems and MD switching for 
larger systems. In the MD regime, finite-size effects on the dynamic order parameter 
$\langle Q \rangle$ and its fluctuations become essentially negligible. This indicates that the 
observed fluctuation peaks are not caused by a critical phenomenon. 
Rather,  they are a stochastic-resonance phenomenon due to the asymmetry 
introduced by the bias field.

The main difference between the behaviors of the two models is that the zero spins 
in the BC model tend to accumulate at the interfaces between regions of positive and 
negative spins. This appears to lead to a reduction of the interface tension between regions of 
positive and negative spins in the BC model, compared with the Ising model.

Future studies may reveal further, novel effects as the dynamics of spin systems with 
multiple states are extended to antiferromagnetic interactions, 
more complicated lattices, and higher spatial dimensions.  

\section*{Acknowledgments}

G.M.B. expresses her appreciation for support and hospitality
at the PoreLab and NJORD Centres of the Department of Physics at the University of Oslo.

Work at the University of Oslo was supported by the Research Council of Norway through the Center of Excellence funding scheme, Project No. 262644.

\section*{Data Availability Statement}
Data will be made available on reasonable request. 

\section*{Author contributions}
All authors contributed equally to this work. 

%\bibliography{ref}

\clearpage

%%%%%%%%%%%%%%% FIGURES

\end{document}